\documentclass[sn-aps]{sn-jnl}

\usepackage{amsmath}
\usepackage{comment}
\usepackage{bm}
\usepackage{xcolor}
\usepackage{url}
\urldef{\mailsa}\path|jaeyun.moon@cornell.edu|    

\theoremstyle{thmstyleone}%
\newtheorem{theorem}{Theorem}
\newtheorem{proposition}[theorem]{Proposition}%

\theoremstyle{thmstyletwo}%
\newtheorem{example}{Example}%
\newtheorem{remark}{Remark}%

\theoremstyle{thmstylethree}%
\newtheorem{definition}{Definition}%

\begin{document}

\title[Heat Carriers in Liquids: an introduction]{Chapter 2:
Normal mode decomposition of atomic motion in solids}

\author[*]{Jaeyun Moon}%

\affil[1]{\orgdiv{Sibley School of Mechanicaland Aerospace Engineering}, \orgname{Cornell University}, \orgaddress{\city{Ithaca}, \postcode{14853}, \state{New York}, \country{USA}}
\mailsa}


\abstract{Decomposition of atomic motion into individual normal modes has led to remarkable success in microscopically understanding thermal properties and thermodynamics in simple solids. We start this chapter with an example of decomposing atomic motion of a simple monatomic linear chain crystal into normal modes followed by a more general, classical normal mode formalism. Different classifications of normal modes such as phonons, propagons, diffusons, and locons are introduced. Finally, heat capacity and thermal conductivity predictions from the normal mode formalism are demonstrated.}




\maketitle

\section{Introduction}\label{sec1}


Characterization of atomic degrees of freedom is crucial in describing and understanding various materials properties microscopically. In simple, perfect solids, atoms vibrate around their equilibrium positions. Normal mode (sinusoidal planewave) decomposition of atomic motion in solids is, therefore, a natural way of understanding the atomic degrees of freedom.  

In this chapter, we describe succinctly what normal modes are and how they can be used to describe some materials properties such as heat capacity and thermal conductivity in solids. Some historically significant and simple models such as Debye \cite{debye_zur_1912} and Einstein \cite{einstein_plancksche_1907} models are discussed. We will primarily focus on normal modes in crystals in this chapter but how the nature of normal modes changes going from crystals to amorphous solids is briefly discussed.

\section{What are normal modes?}
\subsection{Example: monatomic linear chain crystal}
\begin{figure}%
\centering
\includegraphics[width=0.9\textwidth]{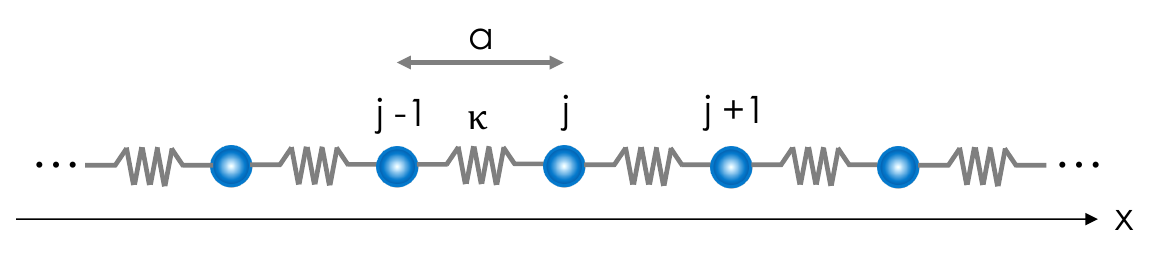}
\caption{Schematic of a linear atomic chain crystal. $\kappa$ is the harmonic force constant or the spring constant and interatomic distance is given by $a$.}\label{fig:monatomic_chain}
\end{figure}

To get a physical intuition for what normal modes are, we first begin by a simple textbook example of describing the equation of motion of a monatomic linear chain crystal in terms of normal modes as shown in Fig. \ref{fig:monatomic_chain}. We assume that interatomic interaction is described by a simple spring motion, ignoring higher order terms in the potential. This treatment of the potential is also known as the \textit{harmonic approximation}. \textit{Anharmonic} terms can be included in the potential for more accurate descriptions of interatomic interactions, often by perturbation methods (described in Chapter 3.2.2). Harmonic approximation is useful in that it allows us to obtain many physical characteristics of the system with a relatively small effort.

In this linear chain crystal, each atom has the same mass, $m$. For simplicity, each atom only interacts with the nearest neighbor atoms that are apart by a lattice constant, $a$, by identical springs with a spring constant, $\kappa$. As atoms vibrate, we define displacement of an atom, $j$, from its equilibrium position by 
\begin{equation}
    u_j = x_j - x_j^0
\end{equation}
where $u_j$ is the displacement, $x_j$ is the instantaneous position, and $x_j^0$ is the equilibrium position of atom $j$. Net force on atom $j$ is then given by
\begin{equation}
    F_j = m\frac{d^2u_j}{dt^2}=\kappa(u_{j+1} - u_j) - \kappa(u_j - u_{j-1}).
\end{equation}

As atoms vibrate around their equilibrium positions, it is reasonable to expect that atomic displacements will be a superposition of sinusoidal waves or modes. We then look for a plane-wave solution form for $u_j$ as $u_j = \sum_k A_ke^{-i(\omega_k t - kx)}$ where $\omega_k$ is the angular frequency and $k$ is the wavevector. Rather than a continuous coordinate, $x$ is restricted to the values of $x=ja$ due to the discrete and periodic nature of this atomic system. This also means that we have limits on the values that $k$ can take in this linear chain crystal: $k = \frac{2\pi m}{Na} = \frac{2\pi m}{L}$, where $m$ is an integer, $N$ is the number of atoms, and $L$ is the length of the chain. Discreteness of wavevectors, $k$, merits more discussions. However, in other phases without a periodic structure which are the foci of the book, the meaning of wavevectors becomes more obscure. Hence, we will not delve into this subject. Plugging in the plane-wave solution and simplifying, we obtain
\begin{equation}
    -m\omega_k^2 = \kappa(e^{ika} + e^{-ika} - 2).
\end{equation}
With some algebra, we obtain the following dispersion relations ($\omega = f(k)$) 
\begin{equation}
    \omega_k = \sqrt{\frac{\kappa}{m}} \bigg\lvert \sin  \Big(\frac{ka}{2}\Big) \bigg\rvert.
    \label{Eq:dispersion}
\end{equation}

\begin{figure}%
\centering
\includegraphics[width=0.8\textwidth]{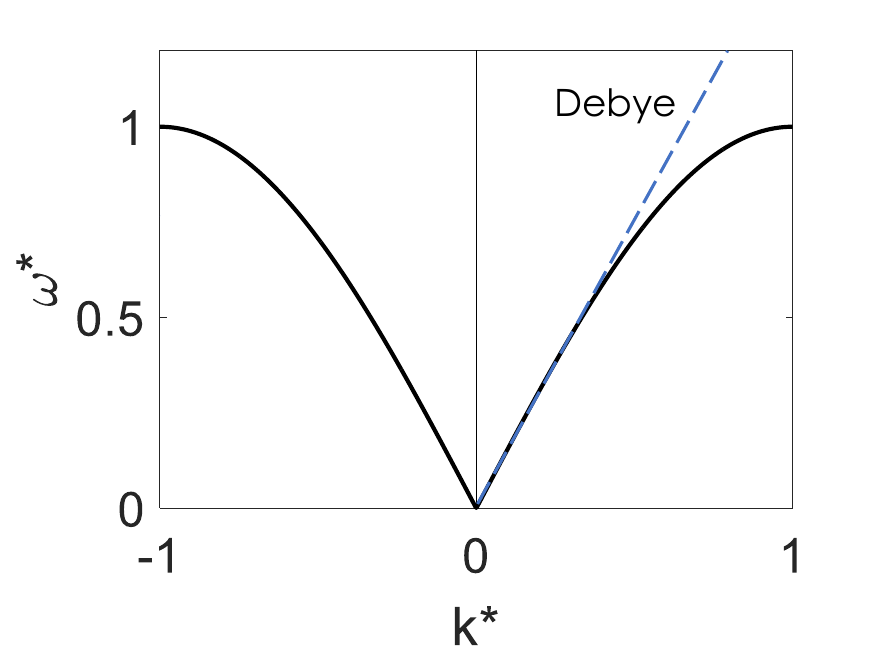}
\caption{Dispersion relations (solid black curves) of a monatomic linear chain crystal as given by Eq. \ref{Eq:dispersion} in reduced units ($k^* = k/(\pi/a)$ and $\omega^* = \omega / \sqrt{\kappa/m}$). At low wavevectors and frequencies, we observe linear dispersions, which are the basis for the Debye model. }\label{fig:linear_dispersion}
\end{figure}

The example given above provides us wealth of information regarding the atomic motion in a crystal. First, dispersion relations shown above demonstrate that atomic motion can be decomposed into individual normal modes with frequencies and wavevectors described by Eq. \ref{Eq:dispersion} under harmonic approximation simply from the structural features (lattice constant, number of atoms, and atomic mass) and force constants. Second, group velocities ($v_g = \frac{d\omega_k}{dk}$) and phase velocities ($v_p = \frac{\omega_k}{k}$) relevant to transport properties and elastic moduli can be found from the dispersion relations. One can, therefore, naturally think that transport properties such as thermal conductivity can be engineered by optimizing some features in the dispersion relations. Third, recognizing that the Taylor expansion of $\sin(x)$ is
\begin{equation}
    \sin(x) = x - \frac{x^3}{3!} + \frac{x^5}{5!} - \frac{x^7}{7!} + \cdot \cdot \cdot,
\end{equation}
the long wavelength limit of the dispersion is 
\begin{equation}
    \omega_k(k \to 0) \sim ck.
\end{equation}
where $c$ is a constant. 
This \textit{linear dispersion} at low wavevectors and low frequencies is observed in various real bulk systems ranging from crystals and glasses to liquids and is the basis for the Debye model \cite{debye_zur_1912}. The Debye model together with Einstein model \cite{einstein_plancksche_1907} will be discussed in detail in Chapter 3.1.1. In this linear dispersion, group velocity and phase velocity are identical to each other at $c$, also known as the sound velocity.

\subsection{Formal description of normal modes in solids}
In the above example, we examined a simple linear chain crystal with one atom per unit cell and its single dispersion curve. More generally, we consider here an arbitrary number of atoms per unit cell with different atomic masses and in three dimensions, which lead to several dispersion curves or \textit{branches} for a given wavevector. In three dimensions with $n$ atoms per unit cell, there are $3n$ branches, of which three are acoustic branches (longitudinal and two transverse) and $3n-3$ are optical branches. Acoustic branches are related to sound waves in solids as demonstrated in the linear chain example and $\omega$ goes to zero as $\boldsymbol{k}$ goes to zero. On the other hand, $\omega$ is non-zero as $\boldsymbol{k}$ goes to 0 for optical modes. The name optical comes about as optical phonons are easily excited by light. 

The equation of motion for atom $j$ in the unit cell $p$ under harmonic approximation is given by
\begin{equation}
    m_j\ddot{\boldsymbol{u}}_{j,p}(t)=-\sum_{j',p'}\boldsymbol{\Phi}_{j,p;j',p'} \cdot \boldsymbol{u}_{j',p'}(t)
    \label{Eq:EOM}
\end{equation}
where matrix elements of the force constants, $\boldsymbol{\Phi}_{j,p;j',p'}$, is given by
\begin{equation}
    \Phi_{j,p;j',p'}^{\alpha \beta} = \frac{\partial^2 U}{\partial u^{\alpha}_{j,p} \partial u^{\beta}_{j',p'}}
\end{equation}
with $\alpha$ and $\beta$ being the Cartesian directions and $U$ being the potential energy. Eq. \ref{Eq:EOM} is essentially a more general case of $F=-\kappa x$. 

As in the linear chain example, we decompose the atomic motion into plane-wave modes as
\begin{equation}
    \boldsymbol{u}_{j,p}(t) = \sum_{\boldsymbol{k},\nu} \boldsymbol{A}(j,\boldsymbol{k}, \nu)e^{i[\boldsymbol{k}\cdot \boldsymbol{r}_{j,p}-\omega(\boldsymbol{k},\nu)t]}
    \label{Eq: PWS}
\end{equation}
where $\boldsymbol{r}_{j,p}$ is the equilibrium position of the atom $j$ in unit cell $p$ and $\boldsymbol{A}(j,\boldsymbol{k}, \nu)$ is the amplitude vector for atom $j$ for mode with wavevector $\boldsymbol{k}$ and branch $\nu$ and is independent of $p$ as the differences in motions between unit cells are described in the exponential phase factor. Plugging in Eq. \ref{Eq: PWS} into Eq. \ref{Eq:EOM} and simplifying, we obtain
\begin{equation}
    m_j\omega^2(\boldsymbol{k}, \nu)\boldsymbol{A}(j,\boldsymbol{k}, \nu) = \sum_{j', p'} \boldsymbol{\Phi}_{j,p;j',p'} \cdot \boldsymbol{A}(j',\boldsymbol{k}, \nu) e^{i\boldsymbol{k}\cdot [\boldsymbol{r}_{j',p'}-\boldsymbol{r}_{j,p}]}.
\end{equation}
In a more compact form, above equation can be re-written as
\begin{equation}
    \boxed{\omega^2(\boldsymbol{k}, \nu) \boldsymbol{e}(\boldsymbol{k}, \nu) = \boldsymbol{D}(\boldsymbol{k}) \cdot \boldsymbol{e}(\boldsymbol{k}, \nu)}
    \label{Eq: eigenvalue problem}
\end{equation}
with the eigenvector $\boldsymbol{e}(\boldsymbol{k}, \nu)$ having $3n$ by 1 matrix elements. $\boldsymbol{e}(\boldsymbol{k}, \nu)$ is related to $\boldsymbol{A}(j,\boldsymbol{k}, \nu)$ by
\begin{equation}
    \boldsymbol{e}(\boldsymbol{k}, \nu) = 
    \begin{bmatrix}
        \sqrt{m_1}A_x(1,\boldsymbol{k}, \nu)\\
        \sqrt{m_1}A_y(1,\boldsymbol{k}, \nu)\\
        \sqrt{m_1}A_z(1,\boldsymbol{k}, \nu)\\
        \sqrt{m_2}A_x(2,\boldsymbol{k}, \nu)\\
        \sqrt{m_2}A_y(2,\boldsymbol{k}, \nu)\\
        \sqrt{m_2}A_z(2,\boldsymbol{k}, \nu)\\
        .\\
        .\\
        .\\
        \sqrt{m_n}A_z(n,\boldsymbol{k}, \nu)
    \end{bmatrix}
\end{equation}
$\boldsymbol{D}(\boldsymbol{k})$ is known as the \textit{dynamical matrix} and is defined as
\begin{equation}
    D^{\alpha \beta}_{jj'}(\boldsymbol{k})=\frac{1}{\sqrt{m_jm_{j'}}}\sum_{p'} \Phi_{j,p;j',p'}^{\alpha \beta}e^{i\boldsymbol{k}\cdot [\boldsymbol{r}_{j'p'}-\boldsymbol{r}_{jp}]}.
    \label{Eq: DM}
\end{equation}
Dynamical matrices can be interpreted as the mass-reduced Fourier transform of the force constant matrices $\Phi_{j,p;j',p'}^{\alpha \beta}$. For each $\boldsymbol{k}$ as shown in Eq. \ref{Eq: DM}, the size of the dynamical matrix is $3n$ by $3n$. The eigenvalue matrix $\Omega(\boldsymbol{k},\nu)=\omega^2(\boldsymbol{k},\nu)$ has the dimensions of $3n$ by $3n$ with only diagonal elements: $\omega^2(\boldsymbol{k},1), \omega^2(\boldsymbol{k},1), \; ... \;, \omega^2(\boldsymbol{k},3n)$. 

Some important characteristics of Eq. \ref{Eq: eigenvalue problem} are worth discussing here. Dynamical matrices are \textit{Hermitian} such that
\begin{equation}
    \boldsymbol{D}(\boldsymbol{k}) = (\boldsymbol{D}^*(\boldsymbol{k}))^T
\end{equation}
and subsequently, eigenvalues, $\Omega(\boldsymbol{k},\nu)$, are always real. This means eigenfrequencies, $\omega(\boldsymbol{k}, \nu)$, can have imaginary and real positive values. Further, corresponding eigenvectors which are generally complex, are orthonormal as
\begin{equation}
    \boldsymbol{e}(\boldsymbol{k}, \nu)^T\cdot \boldsymbol{e}(\boldsymbol{k}, \nu')^* = \boldsymbol{e}(\boldsymbol{k}, \nu)^T\cdot \boldsymbol{e}(\boldsymbol{-k}, \nu') = \delta_{\nu \nu'}
\end{equation}

Nowadays, once we know the structural information and force constants, we can routinely do \textit{lattice dynamics} calculations and solve Eq. \ref{Eq: eigenvalue problem} via widely available softwares such as GULP \cite{gale_gulp:_1997}, LAMMPS \cite{plimpton_fast_1995} and Phonopy \cite{togo_first_2015}.

It is worth mentioning here that primitive unit cells  (smallest unit cell to describe the lattice) are conventionally used in Eq. \ref{Eq: eigenvalue problem} for crystalline materials, but use of other unit cells is equally valid. Different unit cells lead to different dispersion relations from those of primitive unit cells but the spectral distribution function of eigenfrequencies is identical regardless of the choice of the unit cell. For systems with no periodicity (e.g. glasses), the entire domain simulated in computers is considered as a unit cell which means only zero wavevector ($\boldsymbol{k} = \boldsymbol{0}$) dynamical matrices are used \cite{moon_propagating_2018, moon_examining_2021}. When only $\boldsymbol{k} = \boldsymbol{0}$ is considered, we no longer have acoustic or optical branches in a traditional sense as we do not have any other wavevectors. However, algebra becomes more intuitive as dynamical matrices and eigenvectors become real. One can then plot the eigenvector of all atoms for each mode and intuitively understand what individual mode motion looks like. For instance, if a mode is suspected to be spatially localized through some analysis tools, which typically means it contributes negligibly to thermal transport, one can simply visualize the mode and check if only a few atoms are indeed participating in the mode. 

As normal modes describe harmonic atomic motion, we expect that normal modes of amorphous solids which lack long range structural order look or behave differently from those of simple crystalline solids. One example is how far these normal modes can travel before scattering, relevant to thermal conductivity that we will discuss. Imagine waves in the nice calm ocean with some breeze. We expect waves to decay slowly and progressively: these waves can travel long distances similar to phonons in simple crystals. On the other hand, if we have many large and small rocks where the waves are traveling through, the waves will be able to travel only a small distance before scattering. Therefore, we expect normal modes in glasses which are full of defects (like rocks) to have shorter mean free paths (distance before scattering). To discern different types of normal mode characters, prior works have tried to categorize normal modes in solids as phonons, propagons, diffusons, and locons depending on the scattering mechanisms and degrees of structural order. This taxonomy is widely used in studying thermal properties of solids.

\subsection{Types of normal modes in solids: phonons, propagons, diffusons, and locons}

\begin{figure}%
\centering
\includegraphics[width=1\textwidth]{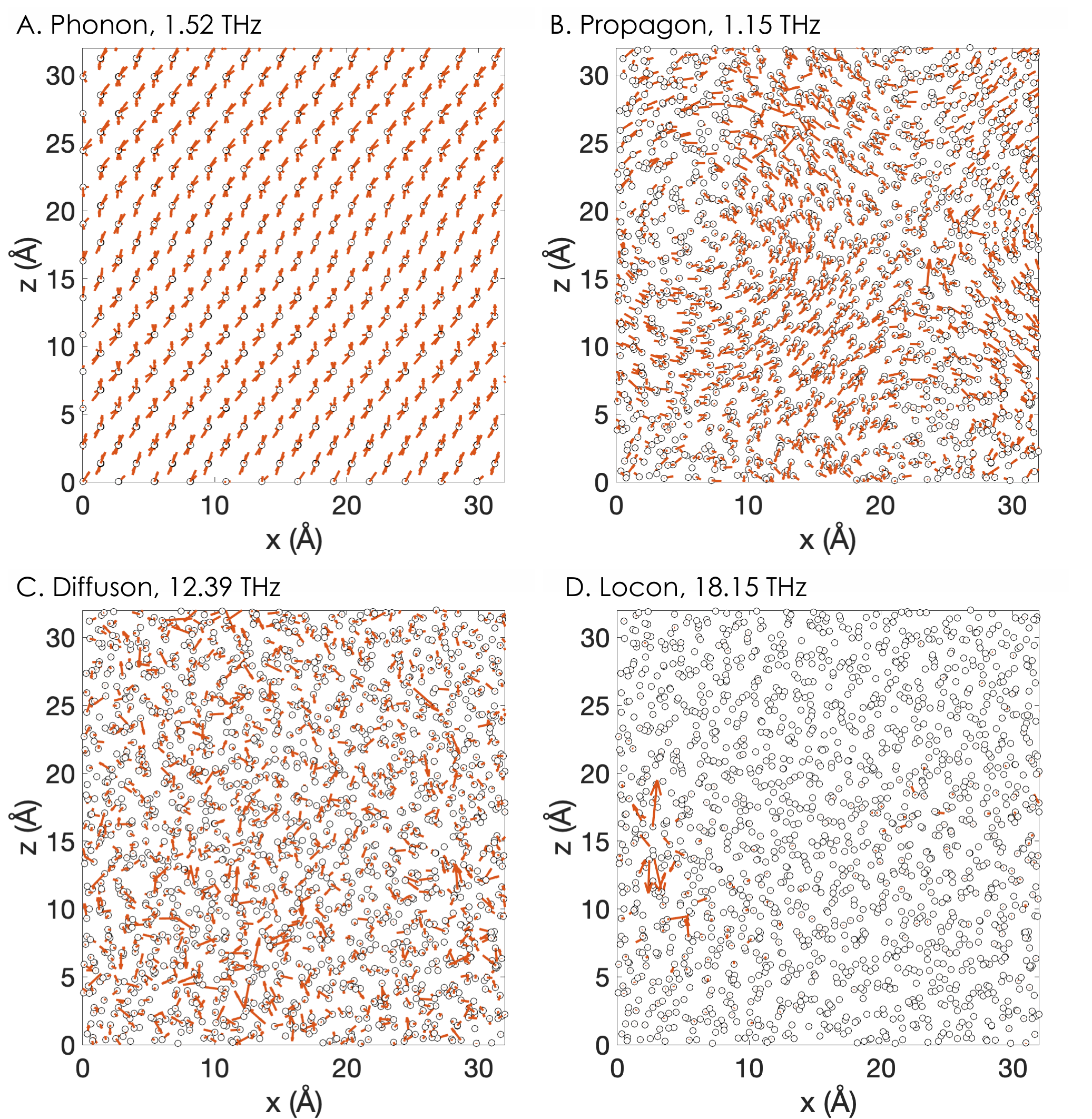}
\caption{2D view (along $y$ direction) of eigenvector of an individual mode (brown arrows) projected on atomic coordinates for (A) acoustic phonon at 1.52 THz for crystalline silicon and (B-D) propagon (1.15 THz), diffuson (12.39 THz), and locon (18.15) for amorphous silicon, respectively. Black circles represent atomic positions. In the normal mode calculations, Stillinger-Weber potential \cite{stillinger_computer_1985} was utilized and both the crystal and glass have the same density. Eigenvector magnitudes are scaled for better visualization. All normal mode calculations were done at $\boldsymbol{k}=0$.}
\label{fig:phonon_locon}
\end{figure}

Phonons, propagons, diffusons, and locons are all vibrational in nature and follow Bose-Einstein statistics. Phonons are quanta of vibrations with well-defined frequencies and wavevectors. As phonon wavevectors are typically associated with spatial periodicity and structural order, phonon terminology is primarily used in crystals. Propagons are similar to phonons in that they are propagating delocalized normal modes that typically possess long wavelengths compared to the interatomic spacing. They too have relatively well-defined wavevectors (sounds also travel to your ears when you place your ears on top of window glasses and tap on them) but not in a traditional sense from periodicity of the structure. Diffusons are modes that scatter over a distance less than interatomic distance and thus transport heat as a random-walk. Locons are non-propagating and localized modes that are very inefficient to transport heat. Propagon, diffuson, and locon terminologies are used primarily when studying disordered systems including glasses \cite{allen_thermal_1989, allen_diffusons_1999, fabian_anharmonic_1996, moon_sub-amorphous_2016, lv_direct_2016, zhou_contribution_2017, agne_minimum_2018, deangelis_thermal_2018}. Representative eigenmode motions for phonons, propagons, diffusons, and locons for a silicon crystal and glass using Stillinger-Weber potential \cite{stillinger_computer_1985} are shown in Fig. \ref{fig:phonon_locon} for a visual aid. Clear changes in the mode behavior are demonstrated.

There are several proposed ways in categorizing normal modes into propagons, diffusons, and locons obtained from diagonalizing dynamical matrix at $\boldsymbol{k} = 0$ by treating the entire domain as a unit cell. There are largely two ways to distinguish propagons and diffusons: transport properties and periodicity of eigenmotion. Regarding the first criterion, prior works have determined the crossover frequency (Ioffe-Regel frequency) when (i) thermal diffusivity of normal modes assuming purely propagon picture or diffuson picture overlaps \cite{larkin_thermal_2014} or (ii) the mean free paths of normal modes become comparable to interatomic distance or their vibrational periods \cite{allen_diffusons_1999, zhu_generalized_2016, he_heat_2011}. Another way to determine the Ioffe-Regel frequency is to use the equilibrium atomic positions and eigenvectors of individual normal modes then look for planewave-like periodicity in the eigenvectors \cite{seyf_method_2016}. The eigenvector periodicity (EP) of a mode is further normalized by a fictitious normal mode that has pure sinusoidal modulation, which provides information about the degree of propagation of that particular mode. Either method typically yields the crossover frequency to be on the order of 1 THz in various amorphous solids. 

For distinguishing diffusons and locons, inverse participation ratio (IPR) has been a widely used metric to find the diffuson to locon crossover frequency (mobility edge) and is also based on the eigenvector characters. In brief, it is a measure of the number of atoms participating in an individual mode. When localized, only a small percentage of atoms will participate in the mode motion as shown in Fig. \ref{fig:phonon_locon}D. On the other hand, if the normal mode is delocalized, there will be many atoms participating in that mode as demonstrated in Fig. \ref{fig:phonon_locon}A-C. 

\begin{figure}
	\centering
	\includegraphics[width=0.8\linewidth]{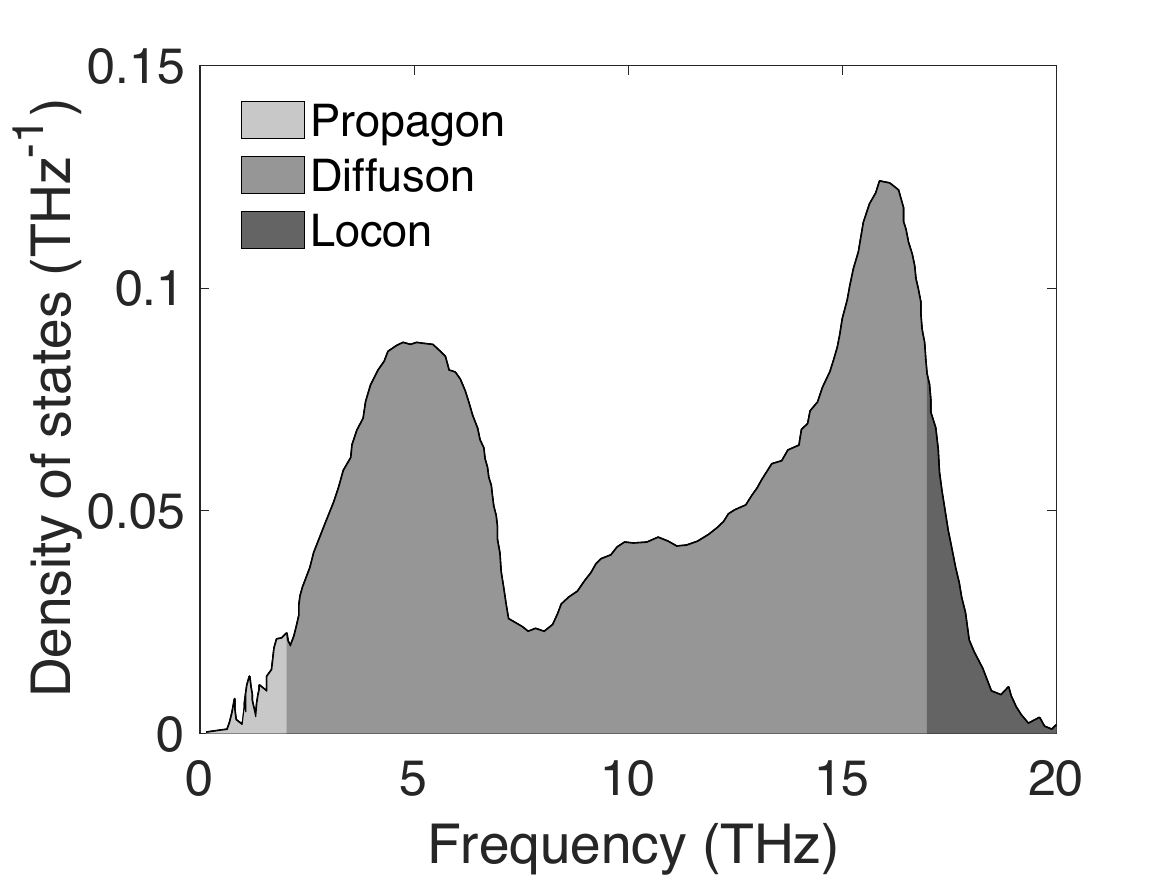}
	\caption{Vibrational density of states of amorphous silicon using Stillinger-Weber potential \cite{stillinger_computer_1985} and 4096-atom size, adapted from  \cite{moon_thermal_2020} Each shaded area represents propagon, diffuson, and locon regions from lighter to darker grey according to Allen and Feldman proposed taxotomy \cite{allen_diffusons_1999}}
	\label{fig:amorphous_DOS}
\end{figure}

Historically, normal modes in amorphous silicon has been studied extensively as a representative amorphous solid due to its simple single-element composition and wide usage in applications such as solar cells \cite{carlson_amorphous_1976} and gravitational wave detectors \cite{birney_amorphous_2018}. A representative normal mode density of states of amorphous solids partitioned into propagons, diffusons, and locons is depicted for amorphous silicon in Fig. \ref{fig:amorphous_DOS}. Most normal modes found in amorphous solids are considered to be diffusons similar to Fig. \ref{fig:amorphous_DOS}. Locons are usually located in the high frequency region of the density of states while some exceptions have also been shown \cite{lv_non-negligible_2016}.

With formal treatment of normal mode decomposition of atomic motion in both ordered and disordered solids, we next discuss understanding thermal properties of solids from normal mode analysis with emphasis on crystals.

\section{Thermal properties from normal modes}
\subsection{Energy and heat capacity}
One important quantity to characterize is the system energy, from which we can obtain important materials properties such as heat capacity and thermal conductivity. For harmonic oscillators, mode energy is given by
\begin{equation}
    E(\boldsymbol{k}, \nu) = \hbar \omega(\boldsymbol{k}, \nu) \Big(n(\boldsymbol{k},\nu) + \frac{1}{2} \Big)
    \label{Eq: harmonic energy}
\end{equation}
where $n(\boldsymbol{k},\nu) = n(\omega,T) = \big [\text{exp}(\hbar \omega(\boldsymbol{k}, \nu)/k_BT) - 1 \big]^{-1}$ is known as the phonon occupation number or Bose-Einstein distribution. System energy under the harmonic approximation is then given by the summation of all the normal mode energy contributions. Individual mode specific heat is defined by $C_V(\boldsymbol{k}, \nu) = \frac{\partial E(\boldsymbol{k}, \nu)}{\partial T} \big \rvert_V$.

\begin{figure}%
\centering
\includegraphics[width=1\textwidth]{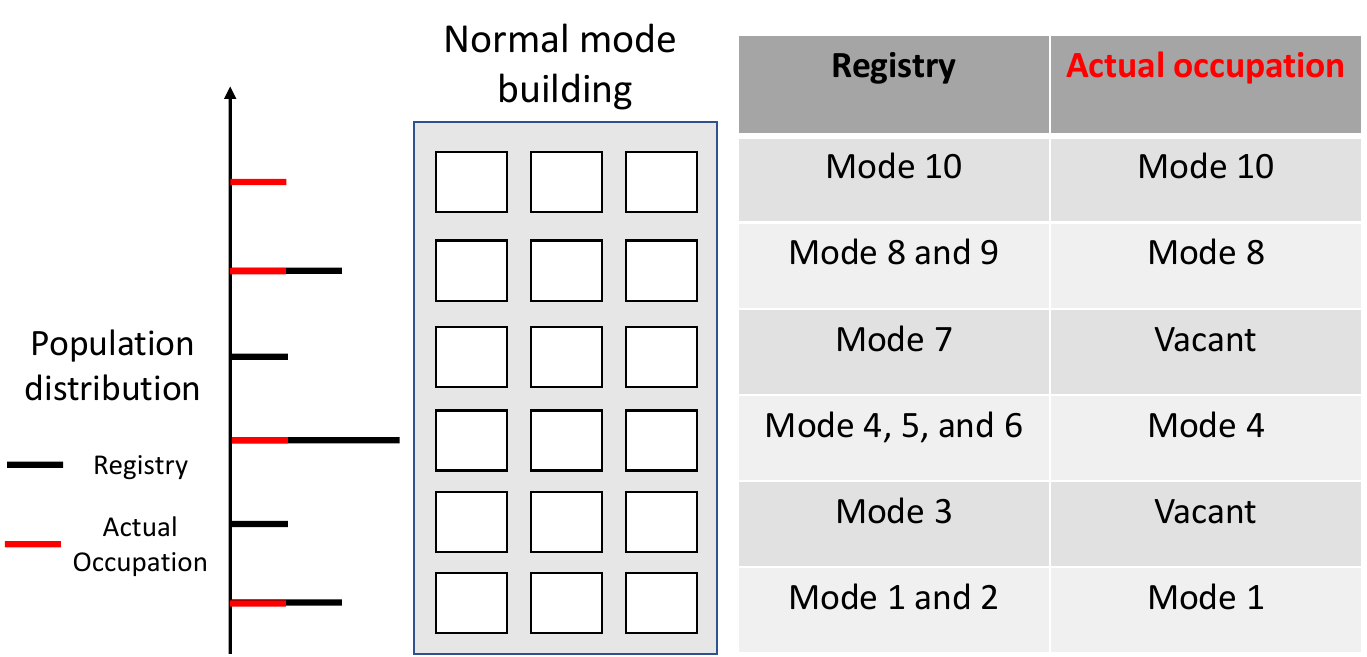}
\caption{Analogical description of density of states in terms of building occupations. Each floor/row of the building represents frequency levels in normal modes. Note that in the actual density of states, frequency levels are not evenly distributed like the building floors shown.}\label{fig:g_schematic}
\end{figure}

As mentioned before in the linear chain example, the meaning of $k$ is not well-defined in systems with no periodicity and it becomes challenging to obtain dispersion relations. It is, therefore, beneficial to develop a formalism that relies only on the frequency distribution. We define here a quantity called the density of states, $g(\omega)$, which describes the population distribution of available normal modes such that it obeys $\int g(\omega) d\omega = 3N$ with $3N$ being the total translational atomic degrees of freedom. A simple analogical description of density of states is shown in Fig. \ref{fig:g_schematic}. Imagine a building where normal modes reside. Floors represent different frequency bins. Building registry is written in the table with some floors with more residents than others. The population distribution of these registered residents (available normal modes) per floor (frequency) is the density of states ($g(\omega))$. However, the actual population of the building can be different from the registry itself as residents may go on vacation, go work during the day, etc. The actual population distribution is a quantum phenomenon and can be described by $g(\omega)n(\omega, T)$. The total energy will be then the energy per floor ($\hbar \omega g(\omega)n(\omega, T))$ integrated by all the floors. More formally, the total system energy in terms of normal modes in frequency is given by
\begin{equation}
     E = \int d\omega \hbar \omega \Big(n(\omega, T)+\frac{1}{2} \Big) g(\omega)  
     \label{Eq: harmonic energy int}
\end{equation}
with addition of $\frac{1}{2}$ arising from the zero point motion. Constant volume heat capacity can then be written as
\begin{equation}
    C_V = \frac{\partial E}{\partial T} \bigg \rvert_V  = \frac{\partial}{\partial T} \int d\omega \hbar \omega \Big(n(\omega, T)+\frac{1}{2} \Big) g(\omega)
    \label{Eq: specific heat}
\end{equation}

The only unknown in Eqs. \ref{Eq: harmonic energy int} and \ref{Eq: specific heat} is the density of states. 

\begin{figure}%
\centering
\includegraphics[width=0.8\textwidth]{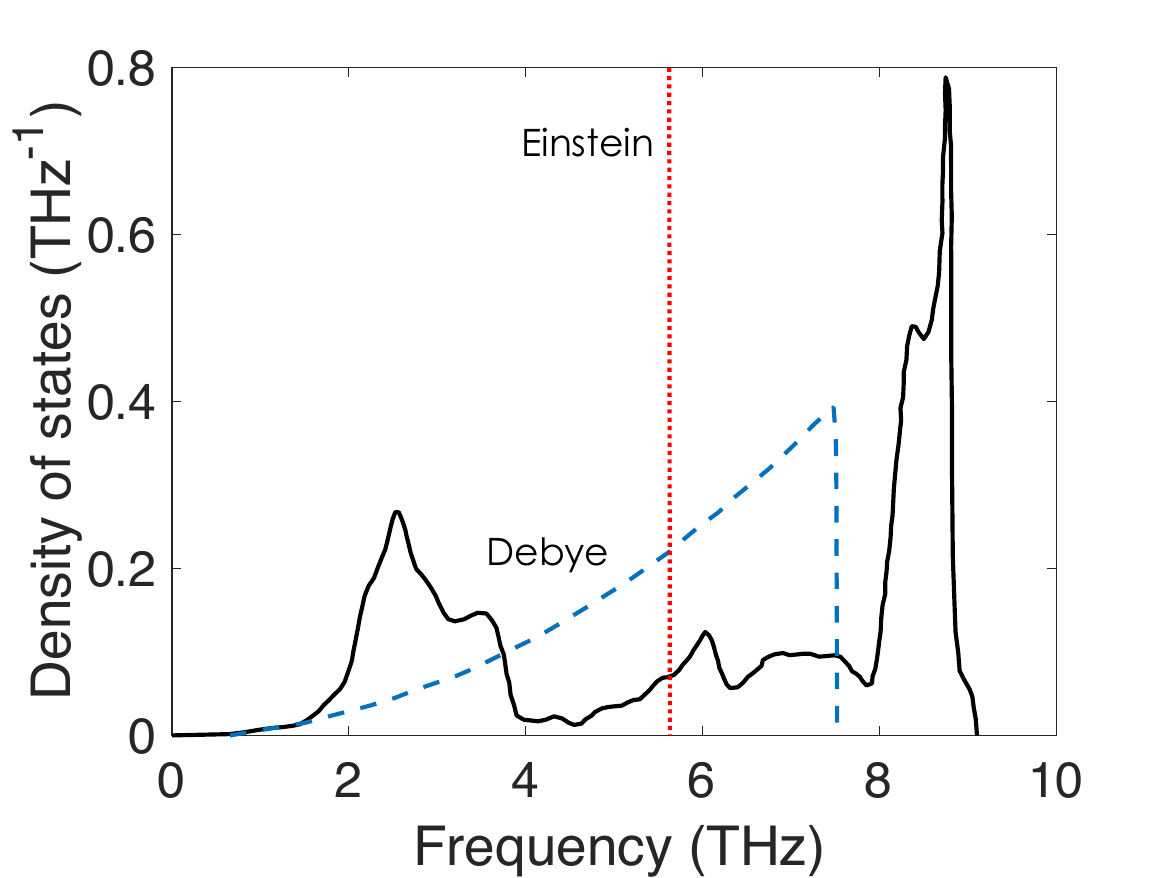}
\caption{Phonon density of states of germanium by density functional theory (solid curves) adopted from Ref. \cite{wei_phonon_1994}. Debye (blue dashed curve) and Einstein (red dotted line) densities of states are also shown for comparisons. In this figure, densities of states are normalized such that the integral over frequency is unity.}
\label{fig:Ge_DOS}
\end{figure}

As mentioned briefly when discussing dispersion relations, density of states can now be routinely found from doing lattice dynamics calculations in readily available softwares. A representative example of phonon density of states of a crystal (germanium) by density functional theory (DFT) \cite{wei_phonon_1994} alongside with Debye \cite{debye_zur_1912} and Einstein \cite{einstein_plancksche_1907} densities of states proposed more than a century ago is shown in Fig. \ref{fig:Ge_DOS}.  

Despite the inaccuracy of Debye and Einstein densities of states in depicting the phonon density of states, they are still used today in complex systems \cite{larkin_thermal_2014, moon_propagating_2018, kim_origin_2021, li_crossover_2019, pocs_giant_2020} for their simplicity, their ability to predict the specific heat of various solids well, and their simple analytical form of density of states. For completeness, we go over these models in detail next.

\subsubsection{Debye and Einstein model}

In the Debye model, we assume a linear dispersion: $\omega = ck$ where $c$ is the averaged sound velocity. In the linear chain example, each wavevector was $2\pi/L$ apart. In the 3D bulk case, each $k$ point has the volume of $(2\pi/L)^3$. Now, consider a sphere of $k$ points. The number of $k$ points with wavevector between $k$ and $k + dk$ is equal to
\begin{equation}
    g(k)dk = \frac{V}{8\pi^3} 4\pi k^2 dk
\end{equation}
where $g(k)$ has the same meaning as $g(\omega)$ but in the wavevector space and $V = L^3$ is the system volume. We can convert this expression from wavevector to frequency space $g(\omega)$ using the linear dispersion and $d\omega = c dk$. We obtain 
\begin{equation}
    g(\omega) d\omega = \frac{3V}{2\pi^2} \frac{\omega^2}{c^3} d\omega
    \label{Eq: Debye_DOS}
\end{equation}
Additional factor of three arises from three polarizations that we need to consider (one longitudinal and two transverse) for 3D bulk systems. Quadratic frequency dependence in the Debye density of states is demonstrated as a parabola in Fig. \ref{fig:Ge_DOS}. In the linear chain example, we discussed that a linear dispersion at low wavevector and frequency is typically found in real systems. This leads to the quadratic frequency dependence in the real phonon densities of states of various materials as demonstrated by the DFT calculated density of states of germanium at low frequencies.  Since the total number of modes is $3N$, there is a restriction of frequency ranges in the Debye density of states from $0$ to $\omega_D$. This upper limit is known as the Debye frequency and is given by
\begin{equation}
    \omega_D = \Big ( \frac{6\pi^2c^3N}{V} \Big )^{\frac{1}{3}}.
\end{equation}
Since $V$ and $c$ are weakly dependent on temperature for crystals, $\omega_D$ is typically assumed as a constant.

Using the Debye density of states, we obtain 
\begin{equation}
    E = \int_0^{\omega_D} \bigg(\frac{3V\hbar\omega^3}{2\pi^2c^3}\bigg) \bigg( \frac{1}{e^{\frac{\hbar \omega}{k_BT}} - 1} +\frac{1}{2} \bigg)  d\omega.
\end{equation}
For simplified expressions, we make the following substitutions
\begin{equation}
    x = \frac{\hbar \omega}{k_B T}
\end{equation}
and rewrite the energy as
\begin{equation}
    E = \bigg ( \frac{3V \hbar}{2\pi^2 c^3} \bigg) \bigg( \frac{k_BT}{\hbar} \bigg )^4 \int_0^{x_D} \frac{x^3}{e^x - 1} dx = 9Nk_BT \bigg(\frac{T_D}{T} \bigg)^3 \int_0^{x_D} \frac{x^3}{e^x - 1} dx,
\end{equation}
where $T_D = \hbar \omega_D/k_B$ is the Debye temperature. Above energy expression can be solved numerically. At low temperatures, $x_D$ can be taken as infinity such that we can make use of 
\begin{equation}
    \int_0^\infty  \frac{x^3}{e^x - 1} dx = \frac{\pi^4}{15},
\end{equation}
which leads to the quartic temperature dependence in energy as $E = \frac{V \pi^2 k_B^4}{10c^3\hbar^3}T^4$ at low temperatures.

Specific heat for the Debye model is then 
\begin{equation}
    C_V = \frac{\partial }{\partial T} \Bigg[ \bigg ( \frac{3V \hbar}{2\pi^2 c^3} \bigg) \bigg( \frac{k_BT}{\hbar} \bigg )^4 \int_0^{x_D} \frac{x^3}{e^x - 1} dx \Bigg ],
\end{equation}
which can be solved numerically. Nonetheless, at two limits of temperatures ($T \to 0$ and $T \to \infty$), analytical forms of the specific heat can be obtained. At low temperatures where we have quartic temperature dependence in energy, we obtain the famous $C_V(T \to 0) \sim T^3$ relations.
In the other high temperature limit, $e^x \sim 1 + x$ such that $C_V(T \to \infty) = 3Nk_B$, also known as the Dulong-Petit law of specific heat.

In contrast to the Debye model where a linear dispersion was asssumed, Einstein postulated that atomic vibrations of a solid were harmonic oscillators, all vibrating at the same frequency, now known as Einstein frequency, $\omega_E$ \cite{einstein_plancksche_1907}. This means the Einstein model for the phonon density of states is given by 
\begin{equation}
    g(\omega) = 3N\delta(\omega - \omega_E)
\end{equation}
as demonstrated as a delta function for germanium in Fig. \ref{fig:Ge_DOS}. The total energy of harmonic oscillators is then 
\begin{equation}
    E = 3N \int_0^\infty \hbar \omega \Big(n(\omega, T)+\frac{1}{2} \Big) \delta(\omega - \omega_E) d\omega
\end{equation}
Taking the temperature derivative, the Einstein model for heat capacity is 
\begin{equation}
    C_V = 3Nk_B\bigg(\frac{\hbar \omega_E}{k_B T} \bigg)^2 \frac{e^{\frac{\hbar \omega_E}{k_BT}}}{\big(e^{\frac{\hbar \omega_E}{k_BT}}-1\big)^2}
    \label{Eq: Einstein Cv}
\end{equation}
Einstein frequency, $\omega_E$, can be obtained by fitting Eq. \ref{Eq: Einstein Cv} to available specific heat data. At low temperatures, Eq. \ref{Eq: Einstein Cv} scales as $C_V \sim e^{-\frac{\hbar \omega_E}{k_B T}}$. At high temperatures, we can also assume here $e^x \sim 1 + x$ such that $C_V(T \to \infty) = 3Nk_B$.

\begin{figure}%
\centering
\includegraphics[width=0.8\textwidth]{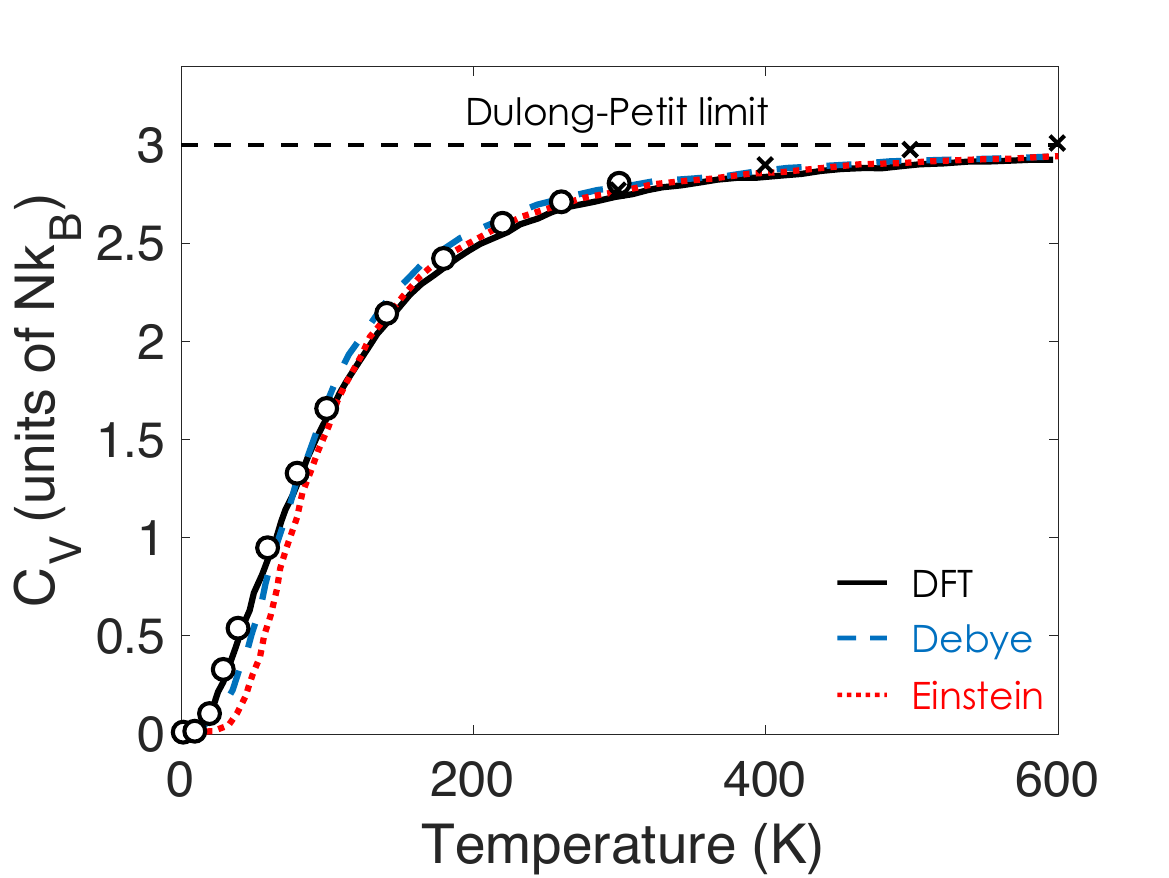}
\caption{Specific heat of germanium. Circles and cross experimental data are from Refs. \cite{berger_semiconductor_1997, lide_crc_2004}. DFT, Debye, and Einstein model predictions of specific heat of germanium are based on the densities of states shown in Fig. \ref{fig:Ge_DOS} and are depicted as solid black, dashed blue, and dotted red curves, respectively. }
\label{fig:Ge_Cv}
\end{figure}

Specific heat predictions of germanium from DFT and Debye and Einstein models are plotted against experimental measurements in Fig. \ref{fig:Ge_Cv}. Despite the simplicity of Debye and Einstein models and their inaccuracy in describing the actual density of states, remarkable agreements are demonstrated throughout all temperatures. At high temperatures, all models reach the Dulong-Petit limit as mentioned previously. At low temperatures, a better agreement between the Debye model and experiments in comparison to the Einstein model is observed. When plotted in the log-log scale (not shown), measurements follow $C_V \sim T^3$ at low temperatures (excluding the electronic contribution) as predicted by the Debye model.  

The theories outlined above for harmonic crystals is applicable under constant volume and harmonic models and cannot describe thermal expansion. However, nearly all experiments are performed under constant pressure rather than constant volume and a simple conversion to the heat capacity at constant volume, $C_P$, can be carried out by
\begin{equation}
    C_P = C_V + \frac{TV\beta^2}{K_T}
\end{equation}
where $\beta$ is the thermal expansion coefficient, $\beta = \frac{1}{V}(\frac{\partial V}{\partial T})_P$ and $K_T$ is the isothermal compressibility, $K_T = -\frac{1}{V}(\frac{\partial V}{\partial T})_T$. $C_V$ and $C_P$ are typically nearly the same at low temperatures but can differ noticeably at high temperatures near melting.

\subsubsection{Some comments about $C_V = 3Nk_B$ at high temperatures}

High temperature limit of $C_V = 3Nk_B$ found both in the Debye and Einstein models is a general result of $N$ classical harmonic oscillators. Each classical harmonic oscillator has the energy of  $E_i = E_{i, KE} + E_{i, PE} = \frac{1}{2}m_i\boldsymbol{v}_i^2 + \frac{1}{2}\kappa \boldsymbol{x}_i^2$ and has three degrees of freedom. From statistical mechanics, each degree of freedom of one harmonic oscillator contributes $k_BT$ to energy ($\frac{1}{2}k_BT$ each from kinetic and potential component of energy), leading to the total energy of $3Nk_BT$. Kinetic and potential component of energy equally contributing to energy is called \textit{equipartition theorem} and derives from having the same quadratic form in energy ($E_{KE} \sim v^2$ and $E_{PE} \sim x^2$). Constant volume specific heat is subsequently $3Nk_B$. 

It is important to use caution when using the result of $C_V = 3Nk_B$ for harmonic oscillators in the high temperature limit. In the high temperature limit calculations  (e.g. thermal conductivity) that utilize the heat capacity of solids, it is often assumed that the constant volume heat capacity is equal to $3Nk_B$ in the literature. However, at very high temperatures, atomic vibration amplitudes can be large enough that atoms can experience higher order potential surfaces even in the constant volume considerations, leading to deviations away from $3Nk_B$ in which case vibrations are not purely harmonic. Assumption of $3Nk_B$ instead of the actual heat capacity leads to difficulties in elucidating mechanisms behind certain features in materials properties convolving heat capacities such as thermal conductivity. Therefore, care must be taken to verify that constant volume heat capacity indeed follows the harmonic oscillator model and $3Nk_B$ at high temperatures.

\subsection{Thermal conductivity}
Another thermal property of crystals that can be described in terms of normal modes is thermal conductivity. Generally, there are several types of heat carriers to consider including electrons and magnons. For dielectric crystals, thermal conductivity is typically dominated by phonons whereas electrons are the major contributors to the total thermal conductivity in metallic systems. Here, we consider phononic contribution to thermal conductivity only. Using simple kinetic theory, we first derive below Fourier's law relating thermal energy current $\boldsymbol{J}$ and thermal conductivity $\boldsymbol{\kappa}$.  

\subsubsection{Fourier's law}

\begin{figure}[h!]%
\centering
\includegraphics[width=1\textwidth]{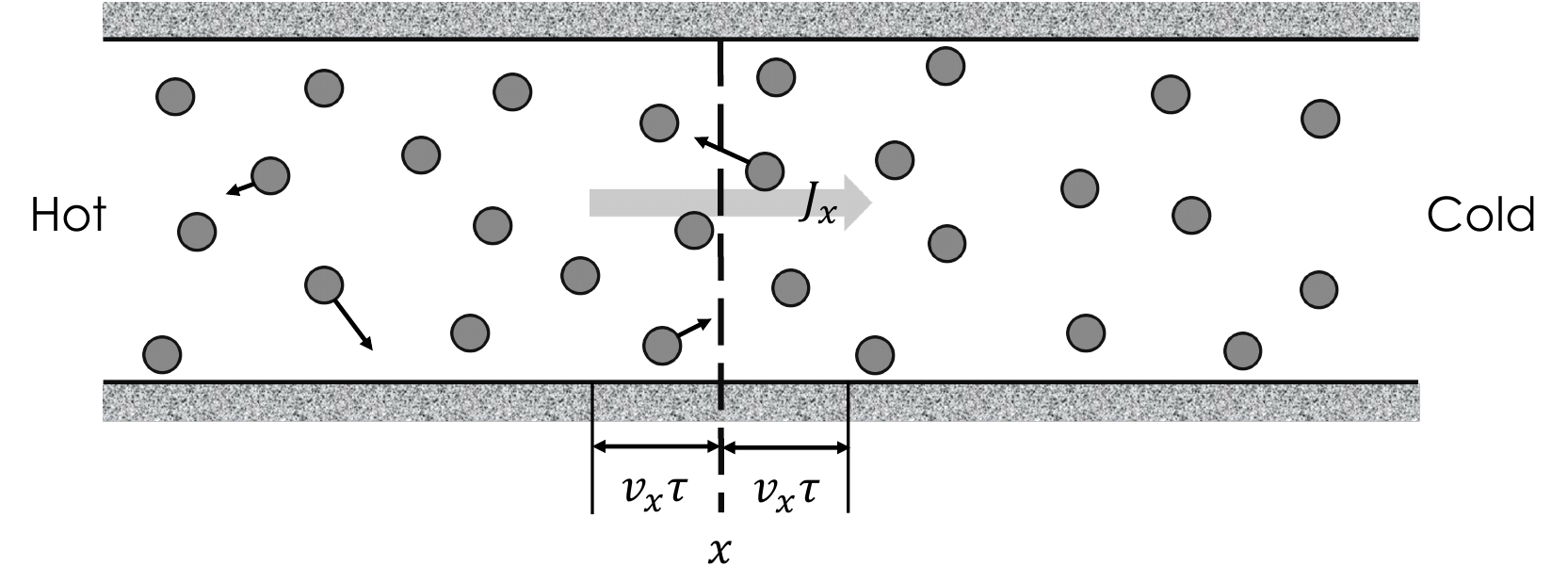}
\caption{Simple derivation of Fourier's law using kinetic theory arguments in one dimension. Heat carriers are represented by grey circles.}
\label{fig:Fourier}
\end{figure}

For simplicity, we only consider one dimension without loss of generality as shown in Fig. \ref{fig:Fourier} where a small overall heat current, $J_x$, flows from hot to cold. If we take an imaginary surface perpendicular to the overall heat flow direction at $x$, the net heat current across this surface is the difference between the energy flow associated with all the heat carriers flowing in the upstream and downstream. Heat carriers within a distance $v_x \tau$ can go across the imaginary surface before being scattered. Here $v_x$ is the $x$ component of the random velocity of the random velocity of the heat carriers and $\tau$ is the relaxation time. $v_x \tau$ is then the distance a heat carrier travels before it is scattered and changes its direction called mean free path, denoted by $\lambda$.  

The net heat current carried by heat carriers across the surface is then
\begin{equation}
    J_x = \frac{1}{2}(nEv_x)\big\rvert_{x-v_x\tau} - \frac{1}{2}(nEv_x)\big\rvert_{x+v_x\tau}
\end{equation}
where $n$ is the number of carriers per unit volume and $E$ is the energy carried. The factor of $1/2$ implies that only half of the heat carriers move in the positive (downstream) direction  while the other half move in the negative (upstream) direction in the presence of small overall heat flow. 

Using Taylor expansion about $x$ up to the first order, we can write the above relation as
\begin{equation}
\begin{split}
        J_x &= \frac{1}{2}(nEv_x)\big \rvert_x + \frac{1}{2}\frac{(dnEv_x)}{dx}(-v_x\tau) - \frac{1}{2}(nEv_x)\big \rvert_x - \frac{1}{2}\frac{(dnEv_x)}{dx}(v_x\tau)\\
        &= -v_x\tau \frac{(nEv_x)}{dx}
\end{split}
\end{equation}

Assuming $v_x$ is independent of $x$ and isotropic such that $v_x^2=(1/3)v^2$, where $v$ is the average random velocity, the above equation becomes
\begin{equation}
    J_x = -\frac{1}{3}v^2\tau \frac{dU}{dT}\frac{dT}{dx}.
\end{equation}
Here $U = nE$ is the local energy density per unit volume. By introducing the heat capacity concept, we obtain the familiar Fourier's law. 
\begin{equation}
    J_x = -\frac{1}{3}(Cv^2\tau)\frac{dT}{dx} = -\kappa \frac{dT}{dx}
\end{equation}
Generally, thermal conductivity can vary spatially and temporally in a material and is a tensor quantity. 

In a more microscopic view, each normal mode (phonon) with wavevector $\boldsymbol{k}$ and on branch $\nu$ carries a portion of the heat current, and overall thermal conductivity can be considered as a summation of all contributions from each normal mode as
\begin{equation}
    \kappa_\alpha = \sum_{\boldsymbol{k}, \nu} \kappa_{\boldsymbol{k}, \nu, \alpha} = \frac{1}{V}\sum_{\boldsymbol{k}, \nu} C_V(\boldsymbol{k}, \nu) v_\alpha(\boldsymbol{k}, \nu)^2 \tau_\alpha(\boldsymbol{k}, \nu)
    \label{eq: thermal conductivity}
\end{equation}
where $V$ is the crystal volume, $C_V(\boldsymbol{k}, \nu)$ is the modal heat capacity as described in section 3.1, $v_\alpha(\boldsymbol{k}, \nu) = d\omega_{\boldsymbol{k},\nu}/dk_\alpha$ is the group velocity in $\alpha$ direction and $\tau_\alpha(\boldsymbol{k}, \nu)$ is the lifetime. 

As discussed previously, the modal specific heat and group velocity can be found from harmonic lattice dynamics calculations. The real challenge in calculating thermal conductivity is accurately determining the phonon lifetimes. In perfectly harmonic solids, there is no loss such that lifetimes are considered to be infinite. In real solids at finite temperatures, atomic motion is not purely harmonic: Bottom of the potential can no longer be approximated as quadratic in spatial coordinates as thermal energy enables atoms to explore more of the potential energy surface. Anharmonicity leads to phonons interacting and having finite lifetimes. 

All the ingredients for thermal conductivity calculations at the mode level then require information about the interatomic interactions whether it is harmonic or anharmonic. For more accurate calculations of the interatomic potential and force constants, density functional theory coupled with perturbation theory is typically utilized at the expense of high computational costs. On the other hand, for materials with low symmetry and large unit cells where density functional theory becomes difficult to utilize, empirical methods such as molecular dynamics can compliment density functional theory in qualitatively characterizing the modal properties needed for thermal conductivity predictions. 

We next discuss the formalisms of obtaining lifetimes from perturbation theory and molecular dynamics. Then, we make comparisons between lifetimes obtained from both formalisms for germanium as an example utilizing the same empirical interatomic potential (Tersoff \cite{tersoff_modeling_1989}). Finally, we consider temperature dependent thermal conductivity predictions utilizing these lifetimes. 

\subsubsection{Phonon lifetime from perturbation theory}
We focus our phonon lifetime using perturbation theory discussions on the lowest order anharmonic term ($V_3$, the 3rd order term in the Taylor expansion of the potential) as it is sufficiently accurate around room temperature for many dielectric materials and is given by
\begin{equation}
    V_3 = \frac{1}{3!}\sum_{j,p}\sum_{j',p'}\sum_{j'',p''}\sum_{\alpha\beta\gamma} \Phi^{\alpha\beta\gamma}_{j,p;j',p';j'',p''}u^{\alpha}_{j,p}u^{\beta}_{j',p'}u^{\gamma}_{j'',p''}.
\end{equation}
where $\Phi^{\alpha\beta\gamma}_{j,p;j',p';j'',p''} = \partial^3 U_{j,p;j',p';j'',p''}/\partial u^{\alpha}_{j,p} \partial u^{\beta}_{j',p'} \partial u^{\gamma}_{j'',p''}$ are third derivatives of the interatomic potential at equilibrium. This third order potential interaction gives rise to three-phonon interactions. 
Recently, efforts have been made to consider up to 4th order anharmonic interactions (4-phonon processes) \cite{feng_quantum_2016, ravichandran_phonon-phonon_2020} and it was found that higher order terms beyond the third order are needed for more accurate lifetime calculations, especially at high temperatures. Differences in thermal conductivity predictions using up to 3rd order or 4th order anharmonic interactions will be briefly discussed in section 3.2.4.

\begin{figure}[h!]%
\centering
\includegraphics[width=1\textwidth]{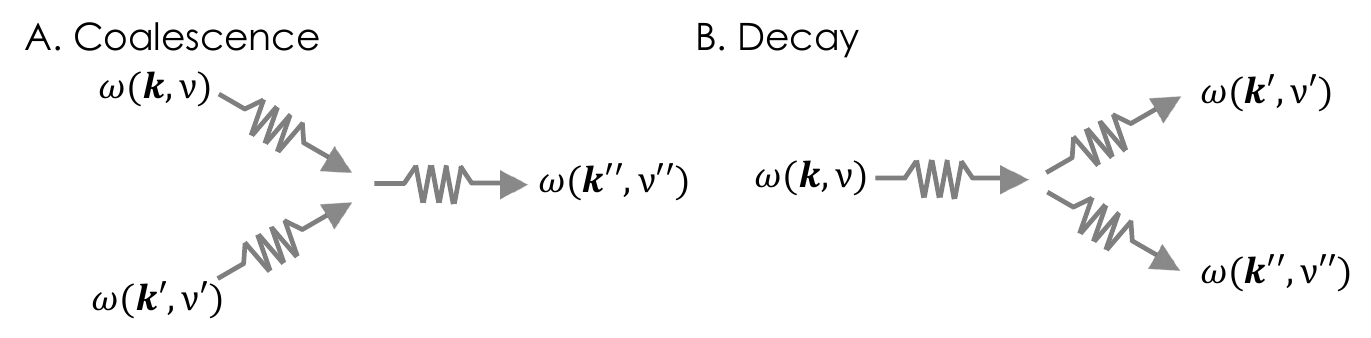}
\caption{3-phonon anharmonic interactions leading to finite lifetimes.}
\label{fig:3phonon}
\end{figure}

The intrinsic phonon lifetimes are given as the sum of all individual scattering probabilities $W = (2\pi/\hbar)\lvert\langle f \lvert V_3 \rvert i \rangle \rvert ^2 \delta(E_f - E_i)$ determined from Fermi's golden rule with the lowest-order anharmonic potential as the perturbation as 
\begin{equation}
    \frac{1}{\tau(\boldsymbol{k}, \nu)} = \sum_{\boldsymbol{k}', \nu'} \sum_{\boldsymbol{k}'', \nu''} \big( W^{\text{coalesce}}_{\boldsymbol{k}, \nu; \boldsymbol{k}', \nu'; \boldsymbol{k}'', \nu''} + \frac{1}{2}W^{\text{decay}}_{\boldsymbol{k}, \nu; \boldsymbol{k}', \nu'; \boldsymbol{k}'', \nu''} \big) 
\end{equation}
\begin{equation}
    \begin{split}
    W^{\text{coalesce}}_{\boldsymbol{k}, \nu; \boldsymbol{k}', \nu'; \boldsymbol{k}'', \nu''} &= \frac{\pi\hbar}{4N}\frac{\lvert \Psi_{-\boldsymbol{k}, \nu; -\boldsymbol{k}', \nu'; \boldsymbol{k}'', \nu''} \rvert^2}{\omega(\boldsymbol{k}, \nu)\omega(\boldsymbol{k}', \nu')\omega(\boldsymbol{k}'', \nu'')} \big[ n(\boldsymbol{k}', \nu') - n(\boldsymbol{k}'', \nu'') \big] \\
    &\quad \delta \big[\omega(\boldsymbol{k}, \nu)+\omega(\boldsymbol{k}', \nu')-\omega(\boldsymbol{k}'', \nu'')] \Delta\big(\boldsymbol{k}+\boldsymbol{k}' - (\boldsymbol{k}'' + \boldsymbol{G})) \\
    \end{split}
\end{equation}
\begin{equation}
    \begin{split}
    W^{\text{decay}}_{\boldsymbol{k}, \nu; \boldsymbol{k}', \nu'; \boldsymbol{k}'', \nu''} &= \frac{\pi\hbar}{4N}\frac{\lvert \Psi_{-\boldsymbol{k}, \nu; \boldsymbol{k}', \nu'; \boldsymbol{k}'', \nu''} \rvert^2}{\omega(\boldsymbol{k}, \nu)\omega(\boldsymbol{k}', \nu')\omega(\boldsymbol{k}'', \nu'')} \big[ n(\boldsymbol{k}', \nu') + n(\boldsymbol{k}'', \nu'') +1 \big] \\
    &\quad \delta \big[\omega(\boldsymbol{k}, \nu)-\omega(\boldsymbol{k}', \nu')-\omega(\boldsymbol{k}'', \nu'')] \Delta\big(\boldsymbol{k}-\boldsymbol{k}' - (\boldsymbol{k}'' + \boldsymbol{G})) \\
    \end{split}
\end{equation}

\begin{equation}
    \begin{split}
   \Psi_{\boldsymbol{k}, \nu; \boldsymbol{k}', \nu'; \boldsymbol{k}'', \nu''} &= \sum_{j}\sum_{j',p'}\sum_{j'',p''}\sum_{\alpha\beta\gamma} \frac{\Phi^{\alpha\beta\gamma}_{j,p;j',p';j'',p''}}{\sqrt{m_jm_{j'}m_{j''}}} \epsilon(\boldsymbol{k}, \nu) ^{\alpha}_{j} \epsilon(\boldsymbol{k}, \nu)^{\beta}_{j'}\epsilon(\boldsymbol{k}, \nu)^{\gamma}_{j''}\\
   &\quad  e^{i\boldsymbol{k}' \cdot \boldsymbol{r}_{j',p'}} e^{i\boldsymbol{k}'' \cdot \boldsymbol{r}_{j'',p''}}
   \end{split}
\end{equation}
for three phonons that conserve energy and momentum within a reciprocal lattice vector $\boldsymbol{G}$:
\begin{equation}
    \omega(\boldsymbol{k}, \nu) \pm \omega(\boldsymbol{k}', \nu') = \omega(\boldsymbol{k}'', \nu'')
\end{equation}
\begin{equation}
    \boldsymbol{k} \pm \boldsymbol{k}' = \boldsymbol{k}'' + \boldsymbol{G}
\end{equation}
where $\pm$ corresponds to coalescence and decay processes, respectively, as shown in Fig. \ref{fig:3phonon}.

Extrinsic factors such as defects and electrons can also scatter phonons and give finite phonon lifetimes and should be included in the calculation of a mode's lifetime. One widely used way to combine different scattering mechanisms is to assume that all scattering mechanisms, both intrinsic and extrinsic, are independent processes. Overall phonon lifetime is then given by the Matthiessen rule as
\begin{equation}
    \frac{1}{\tau(\boldsymbol{k}, \nu)} = \sum_i \frac{1}{\tau_i(\boldsymbol{k}, \nu)}
\end{equation}
where subscript $i$ refers to a scattering mechanism. Characterization of phonon lifetimes due to extrinsic factors is discussed in detail in Refs. \cite{hanus_thermal_2021}.

Anharmonic terms (3rd order and higher orders) in the potential can be often accurately described using the density functional theory but can also be determined from empirical potentials. 

\subsubsection{Phonon lifetime from molecular dynamics}

One major difference between density functional theory and molecular dynamics is in describing interatomic interactions. Molecular dynamics is a classical technique as it uses the Newtonian equations of motion to describe the atomic dynamics. Normal modes in an MD simulation, therefore, follow Boltzmann statistics, the high-temperature limit of Bose-Einstein statistics. Qauntum corrections are often applied to thermal conductivity predictions from molecular dynamics when comparing directly with measurements at low temperatures where consideration of Bose-Einstein statistics is important. In contrast to anharmonic lattice dynamics calculations mentioned in section 3.2.2, MD simulations intrinsically include the full anharmonicity in the interatomic potential and all orders of phonon-phonon interactions are included in phonon lifetime calculations. 

The Hamiltonian of a system of harmonic oscillators can be expressed equivalently in both real space and  normal mode space as follows
\begin{equation}
    H = \frac{1}{2} \sum_{jp} m_j \rvert \boldsymbol{\dot u}_{jp}(t)\rvert^2 + \frac{1}{2} \sum_{jp; j'p'} \boldsymbol{u}^T _{jp}(t) \cdot \Phi_{jp, j'p'} \cdot \boldsymbol{u}_{j'p'}(t)
    \label{harmonic_H}
\end{equation}
and
\begin{equation}
    H = \frac{1}{2} \sum_{\boldsymbol{k}, \nu} \dot Q(\boldsymbol{k}, \nu, t) \dot Q(-\boldsymbol{k}, \nu, t) + \frac{1}{2} \sum_{\boldsymbol{k}, \nu} \omega^2 (\boldsymbol{k}, \nu) Q(\boldsymbol{k}, \nu, t) Q(-\boldsymbol{k}, \nu, t)
\end{equation}
where a dot over a variable represents a time derivative, superscript $T$ represent a transpose. $Q(\boldsymbol{k}, \nu, t)$ often referred to as a normal mode coordinate is the Fourier transform of the $\boldsymbol{u}_{jp}(t)$ and is given by 
\begin{equation}
Q(\boldsymbol{k}, \nu, t) = \frac{1}{N^{\frac{1}{2}}} \sum_{jp}m_j^{\frac{1}{2}}e^{-i\boldsymbol{k}\cdot \boldsymbol{r}_{jp}}\boldsymbol{e}^*(j,\boldsymbol{k}, \nu) \cdot \boldsymbol{u}_{jp}(t)
\end{equation}

The first term and the second term in the above Hamiltonian expressions are kinetic energy and potential energy, respectively. The instantaneous, total energy of each mode of a classical system is then
\begin{equation}
    H_{\boldsymbol{k},\nu} =  \frac{1}{2} \dot Q(\boldsymbol{k}, \nu, t) \dot Q(-\boldsymbol{k}, \nu, t) + \frac{1}{2} \omega^2 (\boldsymbol{k}, \nu) Q(\boldsymbol{k}, \nu, t) Q(-\boldsymbol{k}, \nu, t)
\end{equation}
To account for the mode interactions, displacements from molecular dynamics (MD) at a desired finite temperature which account for all degrees of anharmonicities are used instead and projected to the normal mode coordinates. The temporal decay of the autocorrelation of $H_{\boldsymbol{k},\nu}$ is then related to the relaxation time of each mode by \cite{ladd_lattice_1986,mcgaughey_phonon_2006}
\begin{equation}
    \frac{H_{\boldsymbol{k},\nu}(t) H_{\boldsymbol{k},\nu} (0)}{H_{\boldsymbol{k},\nu}(0) H_{\boldsymbol{k},\nu}(0)} = e^{-\frac{t}{\tau_{\boldsymbol{k}, \nu}}}
\end{equation}

\begin{figure}[h!]
	\centering
	\includegraphics[width=0.72\linewidth]{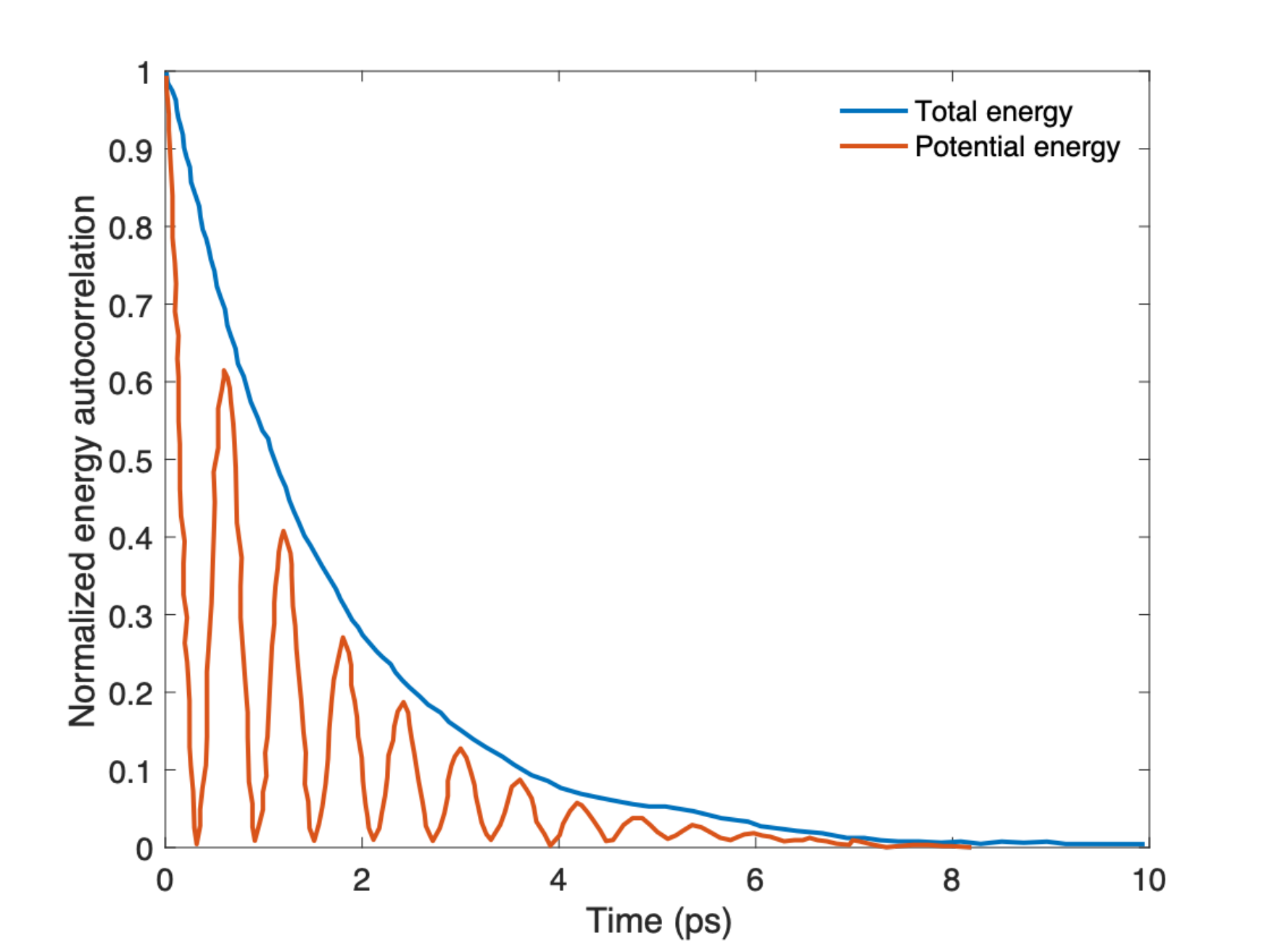}
	\caption{A typical normalized energy autocorrelation for the relaxation time calculations as adapted from \cite{mcgaughey_phonon_2006}. Blue and orange solid curves correspond to normalized total energy and potential energy autocorrelations of a mode, respectively. We see an exponential decay in the total energy autocorrelation as expected. The vibration frequency of the mode is one half of the oscillation frequency observed in the potential energy autocorrelations.  }
	\label{fig:lifetime_mcgaughey}
\end{figure}

A typical normalized total energy autocorrelation for a mode is shown in Fig. \ref{fig:lifetime_mcgaughey} \cite{mcgaughey_phonon_2006, moon_examining_2021}. As expected, we see an exponential decay in the normalized energy autocorrelation and the lifetime can be extracted as discussed. Normalized potential energy autocorrelation is also plotted in the same figure. The vibration frequency of the mode is half of the oscillation frequency in the potential energy autocorrelation. The normal mode lifetime calculations can also be done in frequency space in which we see a Lorentzian function that determines the mode frequency (peak location) and mode lifetime (inverse of full width at half maximum). 

\subsubsection{Phonon lifetime and thermal conductivity prediction example: germanium}
\begin{figure} [h!]
	\centering
	\includegraphics[width=1\linewidth]{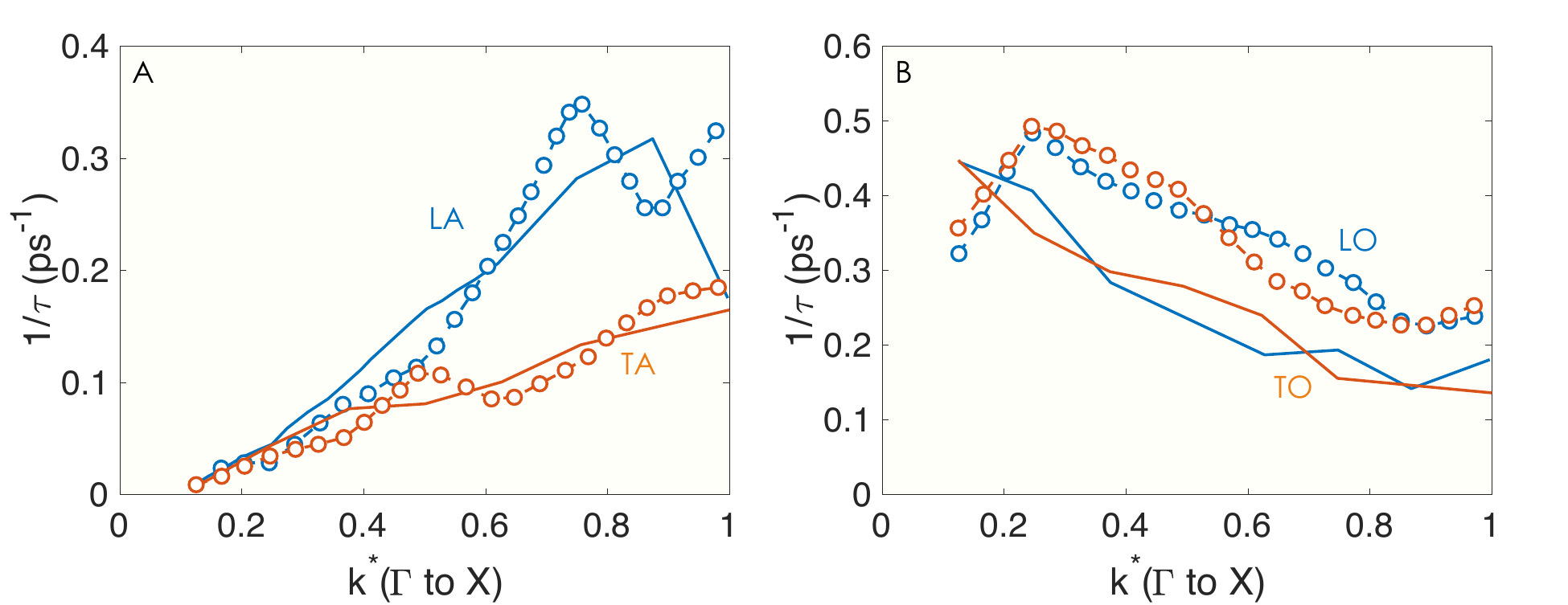}
	\caption{Phonon lifetime comparisons between perturbation theory (solid curves) and molecular dynamics (circles) formalisms described in this Chapter for Germanium at 800 K (adapted from Ref. \cite{feng_quantum_2016}). Perturbation theory here considers up to 4th order term in the potential. Tersoff interatomic potential \cite{tersoff_modeling_1989} is used. Acoustic and optical branch lifetimes along $\boldsymbol{k} = \Gamma$ to $X$ are plotted in (A) and (B), respectively. }
	\label{fig:Ge_lifetime}
\end{figure}

Phonon lifetime comparisons between the perturbation theory and molecular dynamics for certain wavevector directions ($\Gamma$ to X) for Germanium at 800 K is shown in Fig. \ref{fig:Ge_lifetime} \cite{feng_quantum_2016}. For all lifetime calculations, same Tersoff potential \cite{tersoff_modeling_1989} was utilized. Up to 4-phonon processes were considered. For longitudinal and transverse branches (LA and TA), we observe good agreements between lifetime predictions using both methods as shown in Fig. \ref{fig:Ge_lifetime}A. There are some small differences in the lifetimes from these two methods for the optical branches (LO and TO) in Fig. \ref{fig:Ge_lifetime}B and consideration of even higher order terms such as the 5th order may be needed. However, group velocities for these optical branches are usually much smaller than acoustic branches such that small differences in optical phonon lifetime predictions do not lead to large errors in overall thermal conductivity predictions. 

To calculate thermal conductivity, Eq. \ref{eq: thermal conductivity} is used with the lifetimes from the perturbation theory and molecular dynamics, known as the relaxation time approximation (RTA). The main simplifications made by the relaxation time approximation are that all scatterings destroy phonon momentum and relax the phonon mode into equilibrium Bose-Einstein distribution, and the process is independent of the distribution of other modes. More precisely, the deviation from equilibrium from the temperature gradient of the mode at $\boldsymbol{k}$ and $\nu$ does not depend on the distributions of the other phonon modes that it interacts with. Therefore, equilibrium Bose-Einstein distributions are given for phonons with $\boldsymbol{k}'$ and $\nu'$ and $\boldsymbol{k}''$ and $\nu''$ rather than the actual non-equilibrium distributions that arise from a temperature gradient. For many materials including germanium, this approximation is valid; however, for some materials including high thermal conductivity materials, this approximation leads to inaccurate thermal conductivity. The failure of RTA is primarily due to the existence of significant momentum-conserving phonon scattering that causes the drift motion of phonons \cite{lee_hydrodynamic_2015, cepellotti_phonon_2015, lee_hydrodynamic_2020}. This phenomenon, called phonon hydrodynamics, is similar to liquid flow and has been predicted and observed in ultrahigh thermal conductivity materials including graphitic materials \cite{huberman_observation_2019, jeong_transient_2021, machida_phonon_2020}.

\begin{figure}
	\centering
	\includegraphics[width=0.8\linewidth]{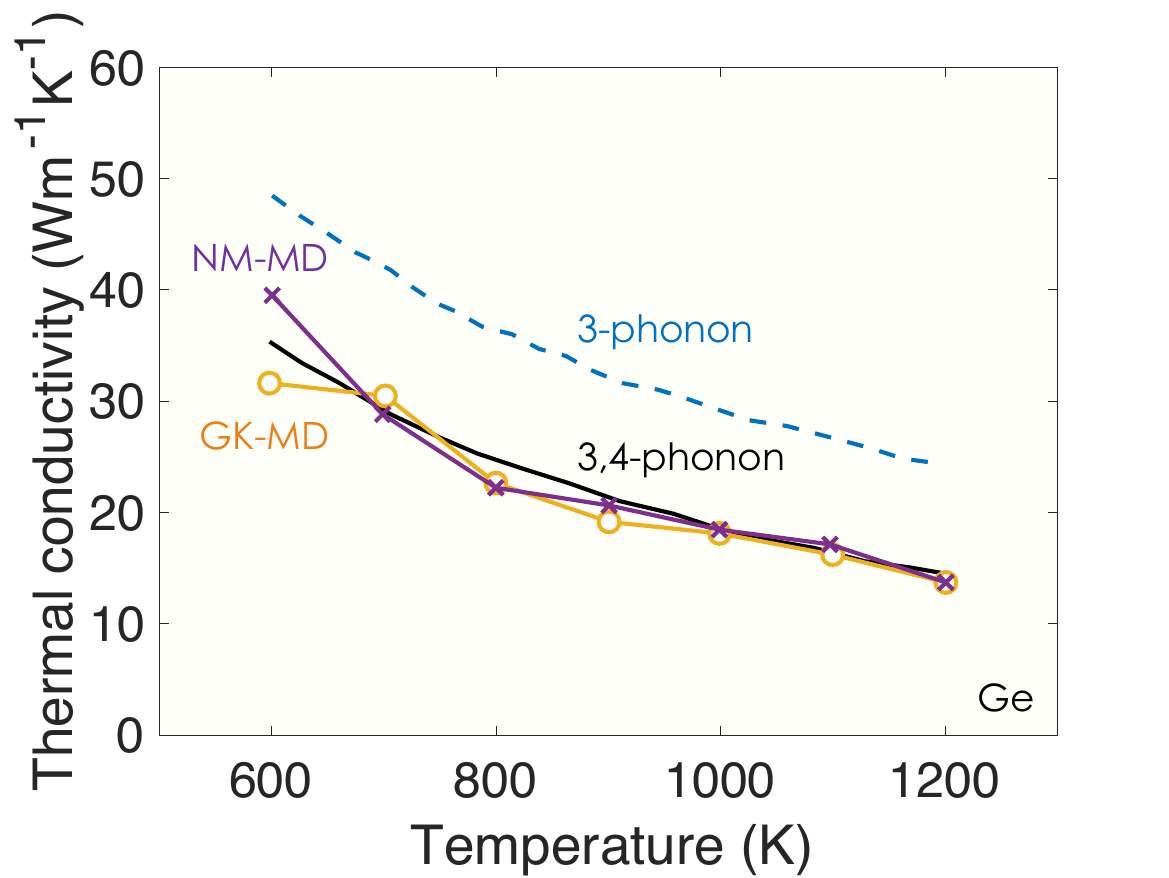}
	\caption{Thermal conductivity predictions of germanium from 3-phonon (blue dashed curve), 3 and 4-phonon (black solid line), and normal mode lifetimes (purple crosses) from molecular dynamics (NM-MD) compared against Green-Kubo (yellow circles) thermal conductivity values (GK-MD) (adapted from Ref. \cite{feng_quantum_2016}).}
	\label{fig:k_Ge}
\end{figure}

Resulting thermal conductivity predictions from the lifetimes obtained from both perturbation theory and molecular dynamics (NM-MD) are compared against thermal conductivity from Green-Kubo formalism (GK-MD) which does not require phonon description of atomic motion and are depicted in Fig. \ref{fig:k_Ge}. Good agreement between perturbation theory (up to 4th order), NM-MD, and GK-MD is demonstrated. Differences between 3-phonon and 3,4-phonon thermal conductivity highlight the importance of considering higher order terms for more accurate thermal conductivity predictions. However, at lower temperatures the differences are typically smaller and at room temperature, difference is less than $\sim$ 5\% \cite{feng_quantum_2016}.

\section{Concluding remarks}

As demonstrated by heat capacity and thermal conductivity discussions, normal mode description of atomic motion in solids has been very successful in describing thermodynamics and various thermal properties of solids. Encouraged by the success, many efforts have been made in recent years to extend the normal mode analysis to liquid systems. Phenomenology of normal mode behaviors and its applications in liquids will be described in Chapter 4.

\bibliography{My_2}



\end{document}